\documentclass[prl,aps,twocolumn,showpacs,amsmath,amssymb]{revtex4}

\usepackage{graphicx}
\usepackage{dcolumn}
\usepackage{bm}

\begin{document}
\title{Different resistivity response to spin density wave and superconductivity at 20 K in $Ca_{1-x}Na_xFe_2As_2$  }
\author{G. Wu, H. Chen, T. Wu, Y. L. Xie, Y. J. Yan, R. H. Liu, X. F. Wang, J. J. Ying,}
\author{ X. H. Chen}
\altaffiliation{Corresponding author} \email{chenxh@ustc.edu.cn}
\affiliation{Hefei National Laboratory for Physical Science at
Microscale and Department of Physics, University of Science and
Technology of China, Hefei, Anhui 230026, P. R. China\\ }
\date{\today}

\begin{abstract}
{We report that intrinsic transport and magnetic properties, and
their anisotropy from high quality single crystal $CaFe_2As_2$.  The
resistivity anisotropy ($\rho_c/\rho_{ab}$) is $\sim 50 $, and less
than 150 of $BaFe_2As_2$, which arises from the strong coupling
along c-axis due to an apparent contraction of about 0.13 nm
compared to $BaFe_2As_2$. Temperature independent anisotropy
indicates that the transport in ab plane and along c-axis direction
shares the same scattering mechanism.  In sharp contrast to the case
of parent compounds $ROFeAs$ (R=rare earth) and $MFe_2As_2$ (M=Ba
and Sr), spin-density-wave (SDW) ordering (or structural transition)
leads to a steep increase of resistivity in $CaFe_2As_2$. Such
different resistivity response to SDW ordering is helpful to
understand the role played by SDW ordering in Fe-based high-$T_c$
superconductors. The susceptibility behavior is very similar to that
observed in single crystal $BaFe_2As_2$. A linear temperature
dependent susceptibility occurs above SDW transition of about 165 K.
Partial substitution of Na for Ca suppresses the SWD ordering
(anomaly in resistivity) and induces occurrence of superconductivity
at $\sim 20$ K. }
\end{abstract}

\pacs{71.27.+a; 71.30.+h; 72.90.+y}

\maketitle
\newpage

The discovery of superconductivity at 26 K in $LaO_{1-x}F_x$FeAs
(x=0.05-0.12)\cite{yoichi}, and $T_c$ surpassing 40 K beyond
McMillan limitation of 39 K predicted by BCS theory in
$RFeAsO_{1-x}F_x$ by replacing La with other trivalent R with
smaller ionic radii \cite{chenxh,chen,ren} have generated much
interest for extensive study on such iron-based superconductors.
which is second family of high-$T_c$ superconductors except for the
high-$T_c$ cuprates. Such Fe-based superconductor shares some
similarity with curpates. They adopt a layered structure with Fe
layers sandwiched by two As layers, each Fe is coordinated by As
tetrahedron. Similar to the cuprates, the Fe-As layer is thought to
be responsible for superconductivity, and R-O layer is carrier
reservoir layer to provide electron carrier. The electron carrier
induced transfers to Fe-As layer to realize superconductivity.
Electronic properties are dominated by the Fe-As triple-layers,
which mostly contribute to the electronic state around Fermi level.

Recently, the ternary iron arsenide $BaFe_2As_2$ shows
superconductivity at 38 K by hole doping with partial substitution
of potassium for barium\cite{rotter}. This material is
$ThCr_2Si_2$-type structure. There exists single FeAs layer in unit
cell in ROFeAs system, while there are two FeAs layers in an unit
cell in $BaFe_2As_2$. These parent compounds share common features:
an anomaly appears in resistivity and such anomaly is associated
with a structure transition and spin density wave (SDW) ordering.
The parent material LaOFeAs shows an anomaly in resistivity at 150 K
which is associated with the structural transition at $\sim 150$ K
and a SDW transition is observed at $\sim 134$ K\cite{yoichi,cruz}.
An anomaly in resistivity occurs at $\sim 140$ K in $BaFe_2As_2$ and
susceptibility shows an antiferromagnetic SDW ordering at almost the
same temperature\cite{wu}. Neutron scattering further indicates that
the antiferromagnetic SDW ordering and structural transition happen
at the same temperature coinciding with the anomaly in
resistivity\cite{huang}. In all systems of $ROFeAs$ (R=rare earth)
and $BaFe_2As_2$ and $SrFe_2As_2$, SDW ordering (or structural
transition) leads to a steep decrease of
resistivity\cite{yoichi,liu,dong,wu,chengf,sasmal}. Such structure
and SDW instabilities are suppressed, and superconducting is induced
and the anomaly in resistivity is completely suppressed by electron-
and hole-doping in $RFeAsO_{1-x}F_x$ system\cite{yoichi,liu,dong}
and $M_{1-x}K_xFe_2As_2$ (M=Ba,Sr)\cite{rotter,chengf,sasmal}. Here
we reported the anisotropy in resistivity and susceptibility in
single crystal CaFe$_2$As$_2$. The resistivity anisotropy is $\sim
50$, and less than 150 of $BaFe_2As_2$\cite{wu}, this is consistent
with an apparent contraction in c-axis lattice relative to
$BaFe_2As_2$. It is striking that the resistivity increases when SDW
ordering occurs in CaFe$_2$As$_2$, in sharp contrast to that SDW
odering leads to steep decrease in resistivity in all other parent
compounds $RFeAsO$ and $MFe_2As_2$ (M=Ba and Sr). Substitution of Na
for Ca leads to suppression of the SDW ordering and structural
transition, and induces superconductivity at $\sim 20$ K. Such
resistivity response to SDW ordering will be helpful to understand
the role played by SDW ordering in Fe-based high-$T_c$
superconductors.

High quality single crystals of CaFe$_2$As$_2$ were grown by
self-flux method, which is similar to that described in our earlier
paper about growth of BaFe$_2$As$_2$ single crystals with FeAs as
flux\cite{wu}. Many shinning plate-like CaFe$_2$As$_2$ crystals were
obtained. The typical dimensional is about 1 x 1 x 0.05 mm$^{3}$.
Polycrystalline samples of Ca$_{1-x}$Na$_x$Fe$_2$As$_2$ was
synthesized by solid state reaction method using CaAs, NaAs and
Fe$_2$As as starting materials. CaAs was presynthesized by heating
Ca lumps and As powder in an evacuated quartz tube at 923 K for 4
hours. NaAs was prepared by reacting Na lumps and As powder at 573 K
for 4 hours, Fe$_2$As was obtained by reacting the mixture of
element powders at 973 K for 4 hours. The raw materials were
accurately weighed according to the stoichiometric ratio of
Ca$_{1-x}$Na$_x$Fe$_2$As$_2$, then the weighed powders were
thoroughly grounded and pressed into pellets. The pellets were
wrapped with Ta foil and sealed in evacuate quartz tubes. The sealed
tubes were heated to 1203 K and annealed for 15 hours. The sample
preparation process except for annealing was carried out in glove
box in which high pure argon atmosphere is filled.

\begin{figure}[t]
\includegraphics[width=9cm]{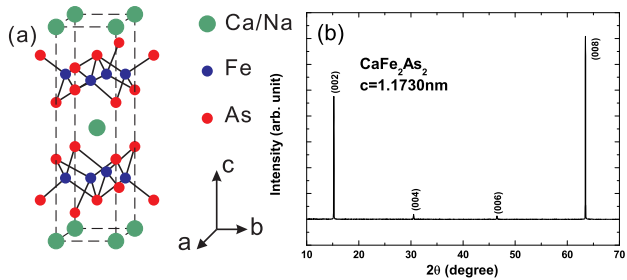}
\caption{ (a): Crystal structure of $CaFe_2As_2$; (b): Single
crystal x-ray diffraction pattern of $CaFe_2As_2$, only (00l)
diffraction peaks show up, suggesting that the c-axis is
perpendicular to the plane of the plate.\\ }
\end{figure}

The crystal structure of Ca$_{1-x}$Na$_x$Fe$_2$As$_2$ is shown in
Fig.1(a), which is the same as BaFe$_2$As$_2$ with the tetragonal
ThCr$_2$Si$_2$-type compound\cite{pfisterer}. The layers of
edge-sharing Fe$_4$As$_4$-tetrahedra are separated by Ca atom
layers. Fig.1(b) shows the single crystal x-ray diffraction pattern
of CaFe$_2$As$_2$. Only (00l) diffraction peaks are observed, it
suggests that the crystallographic c-axis is perpendicular to the
plane of the plate-like single crystal. Table 1 shows the
crystallographic data of single crystal CaFe$_2$As$_2$ at room
temperature. X-ray diffraction intensity data measurement was
performed at 297 K (Mo K$_\alpha$ radiation, $\lambda$=0.71073\AA)
using a Gemini S Ultra (Oxford diffraction). All the structures were
solved by Patterson methods and refined by full-matrix least-squares
methods with SHELX-97\cite{sheldrick}. For comparison, the data of
BaFe$_2$As$_2$ reported by Rotter $et$ $al$\cite{rotter1} are also
listed in Table 1. Both the CaFe$_2$As$_2$ and BaFe$_2$As$_2$ have
the same space group. The lattice parameters of CaFe$_2$As$_2$ is
much smaller than that of BaFe$_2$As$_2$. Compared the average bond
lengths of CaFe$_2$As$_2$ with BaFe$_2$As$_2$, it is found that the
length of Ca$-$As bond is much smaller than Ba$-$As bond. It
indicates that the interaction between the Ca layer and FeAs layer
is much stronger than that between the Ba layer and FeAs layer.

\begin{center}
\begin{table}[tp]%
\caption{Crystallographic data of $CaFe_2As_2$. For comparison, the
data of $BaFe_2As_2$ are also listed, the data are from Ref.13.}
\label{aggiungi} \centering %
\begin{tabular}{clccc}
\toprule %
Temperature=297K     &CaFe$_2$As$_2$   &   &BaFe$_2$As$_2$
\\\hline
   Space group           &I4/mmm  &   &I4/mmm\\\hline
   a (nm)          & 0.3872(9)   &   & 0.39625(1)\\
   b (nm)          & 0.3872(9)   &   & 0.39625(1)\\
   c (nm)          & 1.1730(2)  &   & 1.30168(3)\\
   V (nm$^3$)      & 0.17594(5) &   & 0.20438(1)\\\hline
Atomic parameters:\\
   Ca (Ba)           & 2a (0, 0, 0) &   & 2a (0, 0, 0)\\
   Fe              & 4d ($\frac{1}{2}$, 0, $\frac{1}{4}$) &
    & 4d ($\frac{1}{2}$, 0, $\frac{1}{4}$)\\
   As              & 4e (0, 0, z)  & & 4e (0, 0, z)\\
$ $                & z=0.3665(9)  & & z=0.3545(1)\\\hline
Average Bond lengths (nm):\\
Ca (Ba)$-$As        & 0.3154(0) &   & 0.3379(6)\\
   Fe$-$As          & 0.2370(9) &   & 0.2397(6)\\
   Fe$-$Fe          & 0.2738(7) &   & 0.2799(6)\\\hline
Average Bond angles (deg):\\
   As$-$Fe$-$As$_1$           & 109.5(2) &   & 111.3(6)\\
   As$-$Fe$-$As$_2$           & 109.4(6) &   &
   108.5(6)\\\hline\hline
\end{tabular}
\end{table}
\end{center}

\begin{figure}[b]
\includegraphics[width=9cm]{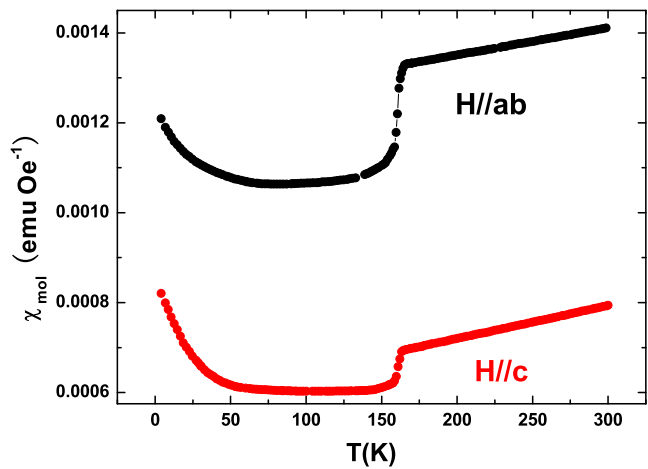}
\caption{Temperature dependence of susceptibility measured under H=5
Tesla applied within ab-plane and along c-axis, respectively, for
single crystal $CaFe_2As_2$.
\\
}
\end{figure}

Temperature dependence of susceptibility measured under magnetic
field of H=5 T applied within ab-plane and along c-axis is shown in
Fig.2, respectively. It should be pointed out that an anisotropy
between H$\parallel ab$ plane and along c-axis is observed, but
these data are not corrected by demagnetization factor.
Susceptibility decreases monotonically for the magnetic field
applied within ab-plane and along c-axis, and shows a linear
temperature dependence above a characteristic temperature of $\sim
165$ K. At 165 K, the susceptibility shows a rapid decrease which is
ascribed to occurrence of antiferromagnetic spin-density wave. Below
165 K, susceptibility decreases more stronger than T-linear
dependence. In low temperatures, a Curie-Weiss-like behavior in
susceptibility is observed. These behaviors are very similar to the
susceptibility behavior reported in single crystal
$BaFe_2As_2$\cite{wu}. The magnitude of susceptibility is almost the
same as that of $BaFe_2As_2$. The susceptibility behavior observed
in both $BaFe_2As_2$ and $CaFe_2As_2$ is very similar to that of
antiferromagnetic SDW pure Cr\cite{fawcett}, in which a temperature
linear dependence persists to the occurrence temperature of SDW.

\begin{figure}[t]
\includegraphics[width=9cm]{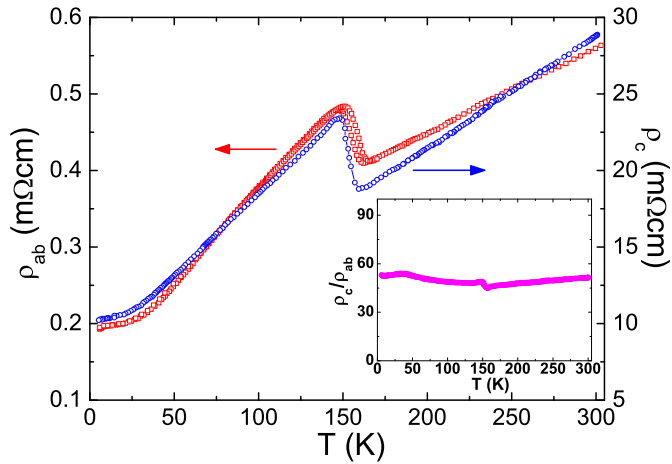}
\caption{Temperature dependence of in-plane and out-of-plane
resistivity ( $\rho_{ab}(T)$ (squares) and $\rho_c(T)$ (circles) )
for single crystal $CaFe_2As_2$. A hysteresis in $\rho_{ab}(T)$ is
observed with cooling and heating measurements. Inset shows
temperature dependence of the anisotropy of resistivity
($\rho_c/\rho_{ab}$). The anisotropy $\rho_c/\rho_{ab}$ is
independent of temperature, indicating that the transport in ab
plane and along c-axis direction shares the same scattering
mechanism. \\}
\end{figure}

Figure 3 shows temperature dependence of in-plane and out-of-plane
resistivity. Both in-plane and out-of-plane resistivity show similar
temperature dependent behavior. In-plane and out-of-plane
resistivities show almost a linear temperature dependence above
$\sim 165$ K, and a steep increase at 165 K, then changes to
metallic behavior. As shown in Fig.3, a hysteresis around 165 K is
observed with cooling and heating measurements, suggesting a
possible first-order transition at 165 K. This transition
temperature coincides with the SDW transition observed in
susceptibility as shown in Fig.2. It indicates that the SDW ordering
leads to a steep increase in resistivity. Such resistivity response
to the SDW transition is in sharp contrast to that for all other
parent compounds $ROFeAs$ (R=rare earth) and $BaFe_2As_2$ and
$SrFe_2As_2$\cite{yoichi,liu,dong,wu,chengf,sasmal}, in which SDW
ordering (or structural transition) leads to a steep decrease of
resistivity. Such different resistivity response to SDW ordering is
helpful to understand what role the SDW ordering plays in Fe-based
high-$T_c$ superconductors. Inset shows the anisotropy of
resistivity ($\rho_c/\rho_{ab}$). The resistivity anisotropy,
$\rho_c/\rho_{ab}$, is about 50. Anisotropy in $CaFe_2As_2$ is less
than 150 of $BaFe_2As_2$. This could arise from that the interaction
between the Ca layer and FeAs layer is much stronger than that
between the Ba layer and FeAs layer. Such strong coupling along
c-axis leads to an apparent contraction of about 0.13 nm in c-axis
lattice, the c-axis lattice parameter decrease from $\sim 1.302$ nm
in $BaFe_2As_2$ to $\sim 1.173$ nm in $CaFe_2As_2$. Almost
temperature independent $\rho_c/\rho_{ab}$ suggests that in-plane
and out-of-plane transports share the same scattering mechanism.

\begin{figure}[t]
\includegraphics[width=9cm]{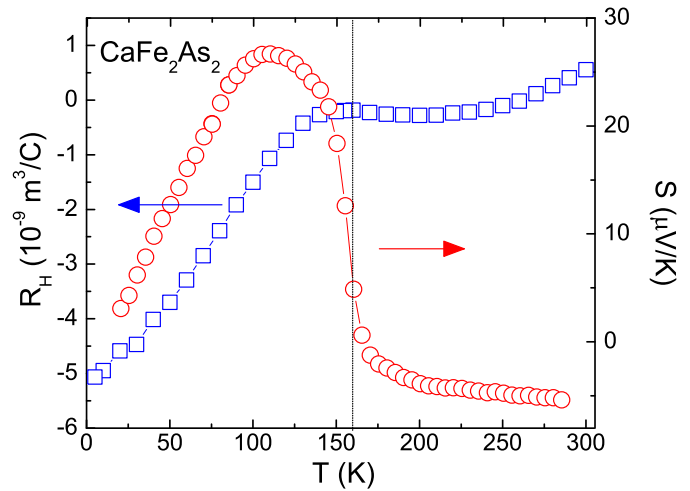}
\caption{The temperature dependence of Hall coefficient and
thermoelectric power (TEP) of single crystal $CaFe_2As_2$.\\}
\end{figure}

Temperature dependences of Hall coefficient and thermoelectric power
(TEP) for single crystal $CaFe_2As_2$ are shown in Fig.4. TEP of
$CaFe_2As_2$ is negative in high temperature range, and changes the
sign at about 165 K and shows a complicated temperature dependence.
TEP slightly increases with decreasing temperature to about 170 K,
and then a big jump increase is observed due to the SDW transition
or structural transition. Below 145 K, TEP slowly increases with
decreasing temperature to about 110 K, then decreases monotonously.
A similar big jump around 140 K was observed in $BaFe_2As_2$, but a
negative sign is opposite to that of $CaFe_2As_2$\cite{wu}. The Hall
coefficient of $CaFe_2As_2$ is negative and changes to positive
above 260 K. The temperature dependence of Hall coefficient is
almost independent on temperature between 260 K and 160 K. Below
$T_s=160$ K, a pronounced decrease in Hall coefficient is observed,
which coincides with $T_s$ of the SDW transition or structural
transition observed in susceptibility and resistivity. The magnitude
of Hall coefficient of $CaFe_2As_2$ at 5 K is about two-order
smaller than that of parent compound of LaOFeAs with single FeAs
layer\cite{McGuire}. It indicates a higher carrier density in
two-layers compounds. The opposite sign of Hall and thermoelectric
power is different with parent compound LaOFeAs\cite{McGuire}, in
which sign of Hall coefficient and thermoelectric power are both
negative. Similar results are also reported in $EuFe_2As_2$
compound\cite{ren}. These results indicate a multi-band scenario in
parent compounds with two FeAs layers in an unit cell.

\begin{figure}[t]
\includegraphics[width=9cm]{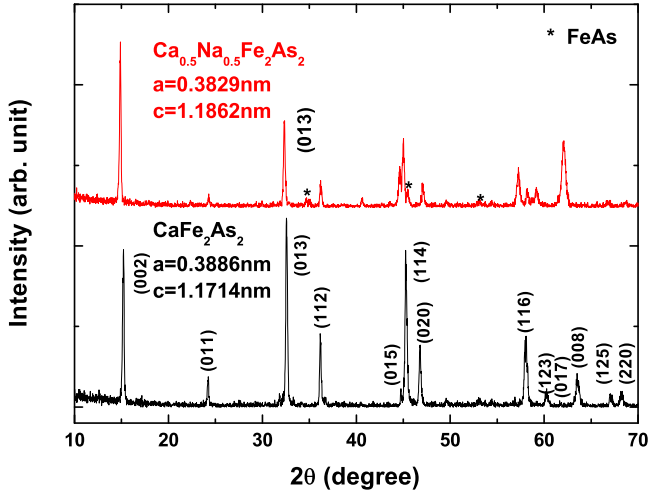}
\caption{X-ray powder diffraction patterns at room temperature for
the polycrystalline samples: CaFe$_2$As$_2$ and
$Ca_{0.5}Na_{0.5}Fe_2As_2$, respectively.\\}
\end{figure}

X-ray powder diffraction patterns are shown in Fig.5 for the
polycrystalline samples: CaFe$_2$As$_2$ and
$Ca_{0.5}Na_{0.5}Fe_2As_2$. Nearly all diffraction peaks in the
patterns of CaFe$_2$As$_2$ and $Ca_{0.5}Na_{0.5}Fe_2As_2$ can be
indexed by the tetragonal ThCr$_2$Si$_2$-type structure, indicating
that the samples are almost single phase. The lattice parameters are
a=0.3886 nm and c=1.1714 nm for the sample CaFe$_2$As$_2$ and
a=0.3829 nm and c=1.1862 nm for the sample
$Ca_{0.5}Na_{0.5}Fe_2As_2$, respectively. It indicates that Na
doping leads to an apparent decrease in a-axis lattice and an
increase in c-axis lattice. This result is similar to
$Ba_{1-x}K_xFe_2As_2$\cite{rotter}.

Figure 6 shows temperature dependence of resistivity for
polycrystalline samples: CaFe$_2$As$_2$ and
$Ca_{0.5}Na_{0.5}Fe_2As_2$. Polycrystalline parent compound shows
similar behavior to that of corresponding single crystal. Similar
temperature-linear dependent resistivity is observed above the
anomaly temperature. The anomaly for increase in resistivity for
polycrystalline sample is much weak relative to that observed in
single crystal. Compared to the single crystal, the transition
temperature in polycrystalline sample is about 10 K higher. As shown
in Fig.6, no anomaly in resistivity is observed
\begin{figure}[t]
\includegraphics[width=9cm]{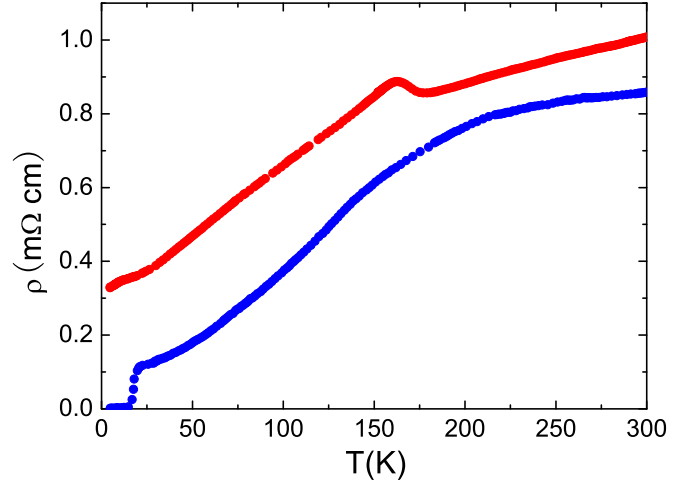}
\caption{Temperature dependence of resistivity for polycrystalline
samples: CaFe$_2$As$_2$ and $Ca_{0.5}Na_{0.5}Fe_2As_2$,
respectively. The anomaly is completely suppressed and a
superconductivity at about 20 K is observed.\\}
\end{figure}
and a superconducting transition at $\sim 20$ K shows up in sodium
doped sample $Ca_{0.5}Na_{0.5}Fe_2As_2$. It suggests that partial
substitution of Na for Ca induces hole-type carrier into system, and
leads to the suppression of structural and SDW instabilities and
induces superconductivity in $Ca_{0.5}Na_{0.5}Fe_2As_2$. Such
behavior is consistent with that reported in electron-doped
$RFeAsO_{1-x}F_x$ (R=rare earth) system\cite{yoichi,liu,ren,dong}
and hole-doped $M_{1-x}K_xFe_2As_2$ (M=Ba and Sr)
system\cite{rotter,wu,chengf,sasmal}.

In summary, we systematically study the anisotropy of resistivity
and susceptibility in high-quality single crystal of parent compound
$CaFe_2As_2$. The resistivity anisotropy ($\rho_c/\rho_{ab}$) is
about 50, and small relative to $BaFe_2As_2$ due to stronger
coupling  between the Ca layer and FeAs layer than that between the
Ba layer and FeAs layer. An apparent contraction along c-axis of
about 0.13 nm is observed. Temperature independent resistivity
indicates that the transport in ab plane and along c-axis direction
shares the same scattering mechanism.  The susceptibility behavior
is very similar to that of antiferroamgnetic SDW pure chromium and
$BaFe_2As_2$. In sharp contrast to the case of other parent
compounds $ROFeAs$ (R=rare earth) and $MFe_2As_2$ (M=Ba and Sr), SDW
ordering (or structural transition) leads to a steep increase of
resistivity. Such different resistivity response to SDW ordering is
helpful to understand the role played by SDW ordering in Fe-based
high-$T_c$ superconductors. Partial substitution of Na for Ca
induces hole-type carrier into system, leading to the suppression of
structural and SDW instabilities and induces superconductivity at
$\sim 20$ K in $Ca_{0.5}Na_{0.5}Fe_2As_2$.

\vspace*{2mm} {\bf Acknowledgment:} This work is supported by the
Nature Science Foundation of China and by the Ministry of Science
and Technology of China (973 project No: 2006CB601001) and by
National Basic Research Program of China (2006CB922005).

\end{document}